\documentclass[numreferences]{kluwer}
\usepackage{savesym}
\savesymbol{iint}
\savesymbol{iiint}
\usepackage{amsmath,latexsym,amssymb,revsymb,verbatim,enumerate,graphicx,klucite}
\restoresymbol{AMS}{iint}
\restoresymbol{AMS}{iiint}

\newtheorem{definition}{Definition}[section]

\newtheorem{lemma}[definition]{Lemma}

\newtheorem{theorem}{Theorem}

\def\nn{\nonumber}
\def\norm#1{ {|\hspace{-.022in}|#1|\hspace{-.022in}|} }
\def\Norm#1{ {\big|\hspace{-.022in}\big| #1 \big|\hspace{-.022in}\big|} }

\def\NORM#1{ {\left|\hspace{-.022in}\left| #1 \right|\hspace{-.022in}\right|} }

\newenvironment{proof}[1][Proof]{\noindent \textbf{{#1~} }}{$\Box$}

\mathchardef\ordinarycolon\mathcode`\:
\mathcode`\:=\string"8000
\def\vcentcolon{\mathrel{\mathop\ordinarycolon}}
\begingroup \catcode`\:=\active
  \lowercase{\endgroup
  \let :\vcentcolon
  }

\newcommand{\nc}{\newcommand}
\nc{\rnc}{\renewcommand} \nc{\beq}{\begin{equation}}
\nc{\eeq}{{\end{equation}}} \nc{\bea}{\begin{eqnarray}}
\nc{\eea}{\end{eqnarray}} \nc{\beqa}{\begin{eqnarray}}
\nc{\eeqa}{\end{eqnarray}} \nc{\lbar}[1]{\overline{#1}}
\nc{\bra}[1]{\langle#1|} \nc{\ket}[1]{|#1\rangle}
\nc{\ketbra}[2]{|#1\rangle\!\langle#2|}
\nc{\braket}[2]{\langle#1|#2\rangle} \nc{\proj}[1]{|
#1\rangle\!\langle #1 |} \nc{\avg}[1]{\langle#1\rangle}

\rnc{\max}{\operatorname{max}} \nc{\rank}{\operatorname{rank}}
\nc{\conv}{\operatorname{conv}}
\nc{\smfrac}[2]{\mbox{$\frac{#1}{#2}$}} 
\DeclareMathOperator{\Tr}{Tr\!}
\DeclareMathOperator{\Var}{Var\!}
\nc{\ox}{\otimes} \nc{\dg}{\dagger} \nc{\dn}{\downarrow}
\nc{\lmax}{\lambda_{\text{max}}}
\nc{\lmin}{\lambda_{\text{min}}}

\nc{\cA}{{\cal A}} \nc{\cB}{{\cal B}} \nc{\cC}{{\cal C}}
\nc{\cD}{{\cal D}} \nc{\cE}{{\cal E}} \nc{\cF}{{\cal F}}
\nc{\cG}{{\cal G}} \nc{\cH}{{\cal H}} \nc{\cI}{{\cal I}}
\nc{\cJ}{{\cal J}} \nc{\cK}{{\cal K}} \nc{\cL}{{\cal L}}
\nc{\cM}{{\cal M}} \nc{\cN}{{\cal N}} \nc{\cO}{{\cal O}}
\nc{\cP}{{\cal P}} \nc{\cQ}{{\cal Q}} \nc{\cR}{{\cal R}} \nc{\cS}{{\cal S}}
\nc{\cT}{{\cal T}} \nc{\cU}{{\cal U}} \nc{\cV}{{\cal V}}
\nc{\cX}{{\cal X}} \nc{\cW}{{\cal W}} \nc{\cZ}{{\cal Z}}

\nc{\CA}{{\cal A}} \nc{\CB}{{\cal B}} \nc{\CC}{{\cal C}}
\nc{\CD}{{\cal D}} \nc{\CE}{{\cal E}} \nc{\CF}{{\cal F}}
\nc{\CG}{{\cal G}} \nc{\CH}{{\cal H}} \nc{\CI}{{\cal I}}
\nc{\CJ}{{\cal J}} \nc{\CK}{{\cal K}} \nc{\CL}{{\cal L}}
\nc{\CM}{{\cal M}} \nc{\CN}{{\cal N}} \nc{\CO}{{\cal O}}
\nc{\CP}{{\cal P}} \nc{\CQ}{{\cal Q}} \nc{\CR}{{\cal R}} \nc{\CS}{{\cal S}}
\nc{\CT}{{\cal T}} \nc{\CU}{{\cal U}} \nc{\CV}{{\cal V}}
\nc{\CX}{{\cal X}} \nc{\CW}{{\cal W}} \nc{\CZ}{{\cal Z}}

\nc{\csupp}{{\operatorname{csupp}}}
\nc{\qsupp}{{\operatorname{qsupp}}} \nc{\var}{\operatorname{var}}
\nc{\rar}{\rightarrow} \nc{\lrar}{\longrightarrow}
\nc{\poly}{\operatorname{poly}}
\nc{\polylog}{\operatorname{polylog}} \nc{\Lip}{\operatorname{Lip}}
\nc{\mb}[1]{\mathbf{#1}}
\nc{\ep}{\epsilon}
\nc{\Om}{\Omega}
\nc{\wt}[1]{\widetilde{#1}}

\def\>{\rangle}
\def\<{\langle}

\def\d{\delta}
\def\e{\epsilon}

\def\s{\sigma}
\def\sig{\sigma}

\def\ph{\varphi}

\def\Ph{\Phi}

\nc{\glneq}{{\raisebox{0.6ex}{$>$}  \hspace*{-1.8ex} \raisebox{-0.6ex}{$<$}}}
\nc{\gleq}{{\raisebox{0.6ex}{$\geq$}\hspace*{-1.8ex} \raisebox{-0.6ex}{$\leq$}}}

\nc{\RR}{{{\mathbb R}}}
\nc{\FF}{{{\mathbb F}}}
\nc{\HH}{{{\mathbb H}}}
\nc{\NN}{{{\mathbb N}}}
\nc{\ZZ}{{{\mathbb Z}}}
\nc{\PP}{{{\mathbb P}}}
\nc{\QQ}{{{\mathbb Q}}}
\nc{\UU}{{{\mathbb U}}}
\nc{\WW}{{{\mathbb W}}}
\nc{\EE}{{{\mathbb E}}}
\rnc{\SS}{{{\mathbb S}}}
\nc{\id}{{\operatorname{id}}}

\nc{\bA}{\mathbb{A}} \nc{\bB}{\mathbb{B}} \nc{\bC}{\mathbb{C}}
\nc{\bD}{\mathbb{D}} \nc{\bE}{\mathbb{E}} \nc{\bF}{\mathbb{F}}
\nc{\bG}{\mathbb{G}} \nc{\bH}{\mathbb{H}} \nc{\bI}{\mathbb{I}}
\nc{\bJ}{\mathbb{J}} \nc{\bK}{\mathbb{K}} \nc{\bL}{\mathbb{L}}
\nc{\bM}{\mathbb{M}} \nc{\bN}{\mathbb{N}} \nc{\bO}{\mathbb{O}}
\nc{\bP}{\mathbb{P}} \nc{\bQ}{\mathbb{Q}} \nc{\bR}{\mathbb{R}} \nc{\bS}{\mathbb{S}}
\nc{\bT}{\mathbb{T}} \nc{\bU}{\mathbb{U}} \nc{\bV}{\mathbb{V}}
\nc{\bX}{\mathbb{X}} \nc{\bW}{\mathbb{W}} \nc{\bZ}{\mathbb{Z}}

\nc{\vholder}[1]{\rule{0pt}{#1}}
\nc{\wh}[1]{\widehat{#1}}
\nc{\h}[1]{\widehat{#1}}

\nc{\ob}[1]{#1}

\def\beq{\begin {equation}}
\def\eeq{\end {equation}}

\nc{\eq}[1]{Eq.~(\ref{eq:#1})} \nc{\eqs}[2]{Eqs.~(\ref{eq:#1}) and
(\ref{eq:#2})}

\nc{\eqn}[1]{Eq.~(\ref{eqn:#1})}
\nc{\eqns}[2]{Eqs.~(\ref{eqn:#1}) and (\ref{eqn:#2})}

\nc{\region}{\cS\cW}

\nc{\fDd}{{\textstyle \frac{D}{d}}}
\nc{\fdD}{{\textstyle \frac{d}{D}}}
\nc{\fDDdd}{{\textstyle \frac{D^2}{d^2}}}
\nc{\fddDD}{{\textstyle \frac{d^2}{D^2}}}
\nc{\fAA}{{\textstyle \frac{|\hspace{-.01in}A_\d\hspace{-.01in}|}{|\hspace{-.01in}A_1\hspace{-.01in}|}}}
\nc{\fAAinv}{{\textstyle \frac{|\hspace{-.01in}A_1\hspace{-.01in}|}{|\hspace{-.01in}A_\d\hspace{-.01in}|}}}
\nc{\fRA}{{\textstyle \frac{|\hspace{-.01in}R\hspace{-.01in}|}{|\hspace{-.01in}A\hspace{-.01in}|}}}
\nc{\fAR}{{\textstyle \frac{|\hspace{-.01in}A\hspace{-.01in}|}{|\hspace{-.01in}R\hspace{-.01in}|}}}
\nc{\fRAd}{{\textstyle \frac{|\hspace{-.01in}R\hspace{-.01in}|}{|\hspace{-.01in}A_\d\hspace{-.01in}|}}}
\nc{\fAdR}{{\textstyle \frac{|\hspace{-.01in}A_\d\hspace{-.01in}|}{|\hspace{-.01in}R\hspace{-.01in}|}}}

\crline{no}

\begin{document}
\begin{opening}
\title{A decoupling approach to the quantum capacity}

\author{Patrick \surname{Hayden}
 \email{patrick@cs.mcgill.ca}}
 \institute{
    School of Computer Science,
    McGill University,
    Montreal, Quebec, Canada
    }

\author{Micha\l{} \surname{Horodecki}
 \email{fizmh@univ.gda.pl}}
 \institute{
   Institute of Theoretical Physics and Astrophysics,
   University of Gd\'{a}nsk, Poland
   }

\author{Andreas \surname{Winter}
 \email{a.j.winter@bris.ac.uk}}
 \institute{
     Department of Mathematics,
    University of Bristol, UK
    }

\author{Jon \surname{Yard}
 \email{jtyard@lanl.gov}}
 \institute{
    CNLS (Center for Nonlinear Studies) \\ 
    CCS-3 (Computer, Computational and Statistical Sciences) \\
     Los Alamos National Laboratories, Los Alamos, NM, USA \\
    Institute for Quantum Information, Caltech, Pasadena, CA, USA
    }

\runningtitle{A decoupling approach to the quantum capacity}
\runningauthor{P.\ Hayden, M.\ Horodecki, A.\ Winter, J.\ Yard}


\begin{abstract}
We give a short proof that the coherent information is an
achievable rate for the transmission of quantum information
through a noisy quantum channel.  Our method is to produce random codes by performing a unitarily covariant projective measurement on a typical subspace of a tensor power state.  We show that, provided the rank of each measurement operator
is sufficiently small, the transmitted data will with high probability be decoupled from the channel's environment.  We also show that our construction leads to random codes whose average input is close to a product state and outline a modification yielding unitarily invariant ensembles of maximally entangled codes.
\end{abstract}

\end{opening}


\section{Background and notation} \label{sec:intro}

There are many seemingly inequivalent operational tasks to perform with many instances of a noisy quantum channel. These range from simulating a noiseless channel on arbitrary inputs to establishing maximal entanglement between the sender and receiver. For a surprising range of such tasks, the optimal achievable rates are the same and are called collectively the quantum capacity of the quantum channel~\cite{S96,bkn,DHW05}. For this paper, we will focus on the latter case, proving the existence of \emph{entanglement generation codes}~\cite{D05} which can create entanglement at any rate less than the coherent information. The result was first conjectured by Schumacher~\cite{S96} and demonstrated with increasing standards of rigor by Lloyd~\cite{L96}, Shor~\cite{S02} and Devetak~\cite{D05}. The proof we give here differs from the previous demonstrations in two ways. First, we construct random codes consisting of states produced by a unitarily covariant measurement on a product state, enabling us to calculate their properties using elementary representation theory, mirroring the approach of \cite{HOW05b} for a related problem. Second, we avoid the need to explicitly construct a decoder for the receiver by reducing the problem of entanglement transmission to decoupling from the channel's environment (see also a recent proof by Klesse~\cite{K07} based on decoupling). Combined, these properties lead to a significantly simplified proof.

To this end, we begin by summarizing our notation and giving enough background to state the coding theorem, assuming a basic background in quantum information theory at the level of \cite{nc}.  We then show that it is sufficient to design codes which remain decoupled from the environment.   In Section~\ref{sec:1shot}, we prove a one-shot version of the coding theorem which, when combined with the results from Section~\ref{sec:iid} on the quantum method of types, gives a proof of the coding theorem for tensor power channels.  In Section~\ref{sec:epilogue} we show how our construction leads to random codes whose average input is close to a product state, and also how a small modification of our method leads to codes which are maximally entangled with a uniformly random subspace of a subspace of the input.   We also comment there on pseudorandom constructions and reflect on the additivity problem for the quantum capacity.

  If $A$ and $B$ are finite dimensional Hilbert spaces, we write their tensor product as $AB \equiv A\otimes B$.  The Hilbert spaces on which linear operators act will be denoted by a superscript.  For instance, we write $\ph^{AB}$ for a density operator on $AB$. Partial traces will be abbreviated by omitting superscripts, such as $\ph^A \equiv \Tr_B\ph^{AB}$.  We use a similar notation for pure states, e.g.\ $\ket{\psi}^{AB}\in AB$, while abbreviating $\psi^{AB} \equiv \proj{\psi}^{AB}$.  The maximally mixed state on $A$ will be written $\pi^A$.  
A generic quantum channel from the density matrices on $A$ to those on $B$ is denoted  $\CN^{A\to B}$.  Note that a linear map $\CN^{A\to B}$ between spaces of operators is a quantum channel if and only if it has a \emph{Stinespring extension}, consisting of a Hilbert space $E$ and an isometry $\CV_\CN^{A\to BE}$ for which $\Tr_E\CV^{A\to BE}_\CN = \CN^{A\to B}$, and that for a given channel such an extension is unique up to isometries on $E$.  Throughout, we will make implicit the action by conjugation of isometries on density matrices, i.e.\ $\CV(\ph) \equiv \CV\ph\CV^\dagger$.
For a bipartite density matrix $\ph^{AB}$, we write
\[H(A)_\ph \equiv H(\ph^A) \equiv -\Tr\ph^A\log\ph^A\]
for the \emph{von Neumann entropy} of $\rho^A$ and take $\log \equiv \log_2$ throughout.  Given a channel $\CN^{A'\to B}$ and any input density matrix $\ph^{A'}$, let $\CV_\CN^{A'\to BE}$ be any Stinespring extension of $\CN^{A'\to B}$. The \emph{coherent information} is defined as
\[I_c(\ph^{A'}, \CN) = H(B)_{\cV(\ph)} - H(E)_{\cV(\ph)}\]
and is independent of the particular Stinespring extension chosen.
A $(Q,n,\e)$
\emph{entanglement generation code} for $\cN$ is a procedure by which a sender and receiver attempt to produce a maximally entangled state 
\[\ket{\Phi}^{R\h{R}} = \frac{1}{\sqrt{2^{nQ}}}\sum_{i=1}^{2^{nQ}}\ket{i}^R\ket{i}^{\h{R}}\]
on two isomorphic Hilbert  $2^{nQ}$-dimensional Hilbert spaces $R$ and $\wh{R}$ in their respective possessions, by using the channel $n$ times. The code consists of an encoding
preparation $\ket{\Psi}^{RA'^n}$ and a decoding channel $\CD^{B^n \rar
\wh{R}}$ satisfying
\begin{equation}
F\Big(\ket{\Ph}^{R\wh{R}},(\openone^R\ox \CD\circ\cN^{\otimes n})(\Psi^{RA'^n})\Big) \geq 1-\e
\label{fidelitycondition}
\end{equation}
where  the \emph{fidelity} \cite{U76} between a pure and a mixed state is defined as
$F\big(\ket{\varphi},\rho\big) = \bra{\varphi}\rho\ket{\varphi}.$
A rate $Q$ is achievable for $\cN$ if there is a
sequence of $(Q,n,\e_n)$ entanglement generation codes for $\cN$
with $\e_n \rar 0$. The \emph{quantum capacity} $\mathcal{Q}(\cN)$ of $\cN$ is the supremum
of the achievable rates.
We will prove the following theorem which gives a lower bound to $\CQ(\CN)$:
\begin{theorem}[Quantum channel coding theorem]
Let $\CN^{A'\to B}$ and $\ph^{A'}$ be given.  Every $0\leq Q < I_c(\ph,\CN)$ is an achievable rate for entanglement generation over $\CN^{A'\to B}$.
\label{theo:maintheorem}
\end{theorem}
A simpler criterion for good codes can be obtained through the following lemma \cite{sw-approx}, allowing us to bypass consideration of the decoding process altogether.  
\begin{lemma}[Sufficiency of decoupling from environment]

Let  $\CV_\CN^{A'\to BE}$ be a Stinespring extension of some channel $\CN^{A'\to B}$.  
Let $\ket{\Psi}^{RA'^n}$ be any encoding with $|R| = 2^{nQ}$, set $\ket{\Psi}^{RB^nE^n} = \CV_\CN^{\ox n}\ket{\Psi}^{RA'^n}$ and let $\ph^{E^n}$ be arbitrary.
Then there exists a decoding map $\CD^{B^n\to \h{R}}$ which, together with the encoding $\ket{\Psi}^{RA'^n}$, comprises a $(Q,n,\ep)$ entanglement generation code for $\CN^{A'\to B}$, provided that  
\begin{equation}
\NORM{\Psi^{RE^n} - \pi^R\otimes \ph^{E^n}}_1 \leq \ep. \label{eqn:th2fidelity} 
\end{equation}
\label{lem:decouple}
\end{lemma}
{\bf Proof:} \, See Appendix.
\\ \\
Here,  the \emph{trace norm} $\norm{X}_1$ of an operator $X$ is the sum of its singular values.

\section{One-shot version} \label{sec:1shot}
This section is devoted to proving a theorem that is at the heart of the proof of the coding theorem for memoryless channels.  Consider a pure state $\ket{\ph}^{AA'}$ and a channel $\CN^{A'\to B}$ with a Stinespring extension $\CV_\CN^{A'\to BE}$ and suppose that $\ph^{A}$ has rank $|A|$.  Let $P$ be a projection onto some subspace $R\subset A$ and let $U$ be a random unitary on $A$ distributed according to Haar measure.  Defining $D = |A|$ and $d = |R|$, the random matrix $V = \sqrt{\fDd} P U$ gives rise to the random unnormalized pure state
\[\ket{\psi}^{RA'} = (V\ox \openone^{A'})\ket{\ph}^{AA'}.\]
If Alice were to send the $A'$ part of the normalized version of this state over the channel, the global state would be the normalized version of $\ket{\psi}^{RBE} = \CV_\CN\ket{\psi}^{RA'}.$
The following theorem  shows that with high probability, the systems $R$ and $E$ are essentially decoupled.  While it is proved in  \cite{HOW05b}, we give a proof here for convenience.
\begin{theorem}[One-shot decoupling theorem]
Let a density matrix $\ph^{AE}$ and a random non-normalized state $\ket{\psi}^{RBE}$ be given as above.  Then
\begin{equation}
\bE
  \Big\| \psi^{RE} - \pi^{R}\ox \ph^E \Big\|_1 
\leq \sqrt{|R| |E| \Tr\,[ (\ph^{AE})^2 ]}.
\label{eqn:oneshot}
\end{equation}
\label{theo:oneshot}
\end{theorem}
In light of the following lemma, this implies a similar bound for the normalized version of $\ket{\psi}^{RBE}$.
\begin{lemma}
For any two density matrices $\rho$, $\sig$ and any $c\in \bR$, 
\[\norm{\rho - \sig}_1\leq 2\norm{c\rho - \sig}_1.\]
\label{lem:trdist}
\end{lemma}
{\bf Proof:}\,\, See Appendix.
\\

\begin{proof}[Proof of Theorem~\ref{theo:oneshot}:]
Our main technique will be to bound the variance of $\psi^{RE}$, by which we mean the trace of its covariance matrix when treated as a random vector.  Indeed, since  $\bE\,\psi^{RE} = \pi^R\ox \ph^E$, we have 
\begin{eqnarray}
\Var\big[\psi^{RE}\big] 
&\equiv&  \bE\NORM{\psi^{RE} - \bE\,\psi^{RE}}_2^2 \nn \\
&=&  \bE\NORM{\psi^{RE} - \pi^{R}\ox \ph^E}_2^2 \nn \\ 
&=& \bE\Tr\Big[\big(\psi^{RE}\big)^2\Big] - \smfrac 1d \bE\Tr\Big[\big(\ph^{E}\big)^2\Big] 
\label{eqn:variance}
\end{eqnarray}
so that by the Cauchy-Schwartz inequality and the concavity of the square-root function, 
\begin{eqnarray*}
\bE \NORM{\psi^{RE} - \pi^{R}\ox \ph^E}_1 &\leq& \bE\sqrt{|R| |E| \Norm{\psi^{RE} - \pi^{R}\ox \ph^E}_2^2} \\
&\leq& \sqrt{|R| |E| \Var\big[\psi^{RE}\big]}. 
  \end{eqnarray*}
Now we write 
 \begin{eqnarray}
\bE\Tr\Big[\big(\psi^{RE}\big)^2\Big] 
&=& \smfrac{D^2}{d^2}\bE\Tr\Big[\big((PU\ox \openone^E)\ph^{AE}(U^\dagger P\ox \openone^E)\big)^2\Big] \nn
\end{eqnarray} 
Let $F_{A}$ be the flip operator acting on two copies of $A$, similarly define $F^{E}$ and abbreviate $F_{AE} = F_A\ox F_E$ and $F_R = (P\ox P)F_A(P\ox P)$.  Because 
\[\Tr\big[F_{AE}(\rho^{AE}\ox \rho^{AE})\big] = \Tr\big[(\rho^{AE})^2\big]\] for any $\rho^{AE}$, cyclicity of the trace and linearity of expectation imply that 
\begin{eqnarray}
\bE\Tr\Big[\big(\psi^{RE}\big)^2\Big] 
&=& \smfrac{D^2}{d^2}\Tr\Big[(\ph^{AE}\ox \ph^{AE}) \big(G\ox \openone^{EE}\big)\Big].
\label{eqn:quadratic}
\end{eqnarray}
where 
\[G = \bE\big[(U^\dagger \ox U^\dagger)  F_R (U\ox U)\big].\] 
We shall now use the following lemma:
\begin{lemma}
\begin{eqnarray*}
G &=& \smfrac 12 \left(\smfrac{d_+}{D_+} + \smfrac{d_-}{D_-}\right) F^{A} + 
 \smfrac 12 \left(\smfrac{d_+}{D_+} - \smfrac{d_-}{D_-}\right){\emph \openone}^{AA}
\end{eqnarray*}
where  $d_\pm = \smfrac{d^2 \pm d}{2}$, $D_\pm = \smfrac{D^2 \pm D}{2}.$
\label{lem:twirl}
\end{lemma}
{\bf Proof:} \,\, See Appendix.
\\ \\
We therefore find that (\ref{eqn:quadratic}) equals
\begin{eqnarray*}
\smfrac 12 \smfrac{D^2}{d^2}\left(\smfrac{d_+}{D_+} + \smfrac{d_-}{D_-}\right)
 \Tr\big[(\ph^{AE})^2\big] + 
\smfrac 12 \smfrac{D^2}{d^2}\left(\smfrac{d_+}{D_+} - \smfrac{d_-}{D_-}\right)
 \Tr\big[(\ph^{E})^2\big].
 \end{eqnarray*}
A straightforward calculation gives the bounds 
\[\smfrac 12 \smfrac{D^2}{d^2}\left(\smfrac{d_+}{D_+} + \smfrac{d_-}{D_-}\right) \leq 1,\,\,\,\,\,
\smfrac 12 \smfrac{D^2}{d^2}\left(\smfrac{d_+}{D_+} - \smfrac{d_-}{D_-}\right) \leq \smfrac 1d\]
so that we obtain 
\begin{eqnarray*}
\bE\Tr\big[(\psi^{RE})^2\big] &\leq&   \Tr\big[(\ph^{AE})^2\big] + \smfrac 1d   \Tr\big[(\ph^{E})^2\big].
\end{eqnarray*}
Because $\ph^{AE}$ and $\ph^B$ have the same spectra, we may combine this bound with (\ref{eqn:variance}) to find that $\Var\big[\psi^{RE}\big] \leq \Tr\big[(\ph^{B})^2\big]$ as required.
\end{proof}

\section{Application to memoryless channels} \label{sec:iid}
In this section, we complete the proof of the coding theorem by
applying the one-shot result (Theorem~\ref{theo:oneshot}) of the
previous section to channels of the form $\CN^{\otimes n}$ obtaining, for any $\ph^{A'}$, codes achieving rates arbitrarily close to $I_c(\ph^{A'},\CN)$.  The
rough idea is that it will be possible to replace the quantities
appearing on the r.h.s.\ of (\ref{eqn:oneshot}) by entropic
quantities because of the memoryless structure of the channel. As a
first step to making this idea precise, we begin by recalling some
needed ideas from the method of types.  
Consider a density matrix with spectral decomposition $\ph^{A'} = \sum_x p_x\proj{x}^A$.  Its $n$'th tensor power can be written as 
\[(\ph^{A'})^{\ox n} = \sum_{x^n}p_{x^n}\proj{x^n}^{A^n}\]
where $p_{x^n} = p_{x_1}\cdots p_{x^n}$ and $\ket{x^n}^{A^n} = \ket{x_1}^A\cdots \ket{x_n}^A$.  The $\d$-(entropy) typical subspace $A_\d\subseteq A^n$ is defined as 
\[A_\d = \text{span}\left\{\ket{x^n}^{A^n} : \left| -\frac 1n \log p_{x^n} - H(\ph^{A'})\right| \leq \d   \right\}\]  
and the \emph{$\d$-typical projection} $\Pi_\d^A$ is defined to project $A^n$ onto $A_\d$.  We shall need the following lemma:
\begin{lemma}[Typicality]
Let a tripartite pure state $\ket{\ph}^{ABC}$ be given.  For every $\delta > 0$ and all sufficiently large $n$ there are $\d$-typical projections $\Pi_\d^{\{A,B,E\}}$ onto $\d$-typical subspaces $A_\d\subseteq A^n$, $B_\d\subseteq B^n$ and $E_\d\subseteq E^n$ such that the states 
\begin{eqnarray}
\ket{\ph}^{A^nB^nE^n} &=& (\ket{\ph}^{ABE})^{\ox n} \label{eqn:typstate1}\\
\ket{\ph_\d}^{A^nB^nE^n} &=& (\Pi_\d^A\ox \Pi_\d^B\ox\Pi_\d^E)\ket{\ph}^{A^nB^nE^n} \label{eqn:typstate2}
\end{eqnarray} 
satisfy 
\begin{eqnarray}
|E_\delta| &\leq& 2^{nH(E)_\ph + n\delta} \label{eqn:typ1}\\
\Tr\Big[\big(\ph_\d^{B_\delta}\big)^2\Big] &\leq& 2^{-nH(B)_\ph + n\delta} \label{eqn:typ2} \\
\Norm{\ph^{A^nB^nE^n} - \ph_\d^{A^nB^nE^n}}_1 &\leq& \ep \label{eqn:typ3}
\end{eqnarray}
where $\ep=2^{-nc\d^2}$ for some constant $c> 0$ independent of $\delta$ and $n$.
\label{theo:typical}
\end{lemma}
{\bf Proof:} \, See \cite{HOW05b}.
\\
\\
We are now ready to prove the quantum channel coding theorem (Theorem~\ref{theo:maintheorem}). As compared to the one-shot theorem (Theorem \ref{theo:oneshot}), which effectively performs a unitarily covariant measurement on the entire purifying space of the input, our codes for $\cN^{\ox n}$ will utilize a measurement on a $\d$-typical subspace $A_\d$ of the entire purifying space $A^n$.  Throughout, we shall appeal to the following monotonicity property of trace distance under the action of a quantum channel (see e.g.\ \cite{nc}):
\begin{eqnarray}
\norm{\ph - \s}_1 &\geq& \Norm{\CN(\ph) - \CN(\s)}_1 \label{trmon}
\end{eqnarray}
\begin{proof}[Proof of Theorem~\ref{theo:maintheorem}:]
Given a channel $\CN^{A'\to B}$ and a density matrix $\ph^{A'}$, fix a Stinespring extension $\CV_\CN^{A'\to BE}$ and a purification $\ket{\ph}^{AA'}$, where $A\simeq A'$.  Feeding part of the purification through the Stinespring extension gives the state $\ket{\ph}^{ABE} = \CV_\CN\ket{\ph}^{AA'}$.  Letting $\delta > 0$ be arbitrarily small, we may invoke Lemma~\ref{theo:typical} to obtain states $\ket{\ph}^{A^nB^nE^n}$, $\ket{\ph_\d}^{A^nB^nE^n}$ and $\d$-typical projections $\Pi_\d^{\{A,B,E\}}$ onto $\d$-typical subspaces $A_\d$,$B_\d$ and $E_\d$ which satisfy (\ref{eqn:typstate1})--(\ref{eqn:typ3}).  Let $P$ be a projection onto some subspace $R\subseteq A_\d\subseteq A^n$ of dimension $|R| = 2^{nQ}$, let $U$ be a Haar random unitary on $A_\d$ and define the random matrix $V = \sqrt{\smfrac{|A_\d|}{|R|}} P U$.  We will show that on average, the normalized version 
\[\ket{\Psi}^{RB^nE^n}=\braket{\psi}{\psi}^{-1/2}\ket{\psi}^{RB^nE^n}\]
of the random unnormalized state
\[\ket{\psi}^{RB^nE^n} = V\ket{\ph}^{A^nB^nE^n}\]
 satisfies 
\begin{equation}
\bE \NORM{\Psi^{RE^n} - \pi^{R}\ox \ph^{E^n}}_1 \leq 6\ep
\label{eqn:finalbound}
\end{equation}
allowing us to conclude the existence of a deterministic encoding for a $(Q,n,6\ep)$ code.  
We begin by defining the random unnormalized state 
\begin{eqnarray*}
\ket{\psi_\d}^{RB^nE^n} &=& V\ket{\ph_\d}^{A^nB^nE^n}.
\end{eqnarray*}
Then by Lemma~\ref{lem:trdist} and the triangle inequality,  
\begin{eqnarray}
 \NORM{\Psi^{RE^n} - \pi^{R}\ox \ph^{E^n}}_1 &\leq&  2\NORM{\psi^{RE^n} - \pi^{R}\ox \ph^{E^n}}_1 \nn  \\
 &\leq& 2\NORM{\psi^{R E^n} - \psi_\d^{RE^n}} +  \label{eqn:est1} \\
 & &  2\NORM{\psi_\d^{RE^n} - \pi^{R}\ox \ph_\d^{E^n}}_1 + \label{eqn:est2}\\
 & &  2\NORM{\pi^{R}\ox \ph_\d^{E^n} -  \pi^{R}\ox \ph^{E^n}}_1 \label{eqn:est3}
\end{eqnarray}
We bound the expectation of (\ref{eqn:est1}) using the following lemma:
\begin{lemma}
Let $V$ be a random linear operator on a finite dimensional space for which $\bE\, V^\dagger V \leq {\emph \openone}$.  Then every Hermitian $X$ satisfies
\[\bE \norm{V XV^\dagger}_1 \leq \norm{X}_1.\]
\label{lem:covariant}
\end{lemma}
{\bf Proof:}\, See Appendix.
\\ \\
Because $\bE \,V^\dagger V = \Pi^A_\delta$,    
Lemma~\ref{lem:covariant} and (\ref{eqn:typ3}) imply
 \begin{eqnarray}
 \bE \NORM{ \psi^{RE^n} - \psi_\d^{RE^n}}_1 &=& \bE\NORM{V(\ph^{A^nE^n} - \ph^{A^nE^n}_\d)V^\dagger}_1 \nn \\
 &\leq& \NORM{ \ph^{A^nE^n} - \ph^{A^nE^n}_\d} \leq \ep.
 \label{eqn:bound1}
 \end{eqnarray}
The third term can be immediately bounded using monotonicity (\ref{trmon}) and the estimate (\ref{eqn:typ3}):
\begin{eqnarray}
\NORM{\pi^{R}\ox \ph_\d^{E^n} -  \pi^{R}\ox \ph^{E^n}}_1 
&=& \NORM{\ph_\d^{E^n} -  \ph^{E^n}}_1 \nn\\
&\leq& \NORM{\ph_\d^{A^nB^nE^n} -  \ph^{A^nB^nE^n}}_1 \leq \ep.
\end{eqnarray}
Now if $0\leq Q < H(B) - H(E) - 3\d = I_c(\ph^{A'},\CN) - 3\d$, 
the expectation of the second term (\ref{eqn:est2}) can be bounded by combining Theorem~\ref{theo:oneshot}, (\ref{eqn:typ1}) and (\ref{eqn:typ2}): 
\begin{eqnarray}
\bE\NORM{\psi_\d^{RE^n} - \pi^{R}\ox \ph_\d^{E^n}}_1 
&\leq& \sqrt{|R||E_\d| \Tr\Big[\big(\ph_\d^{B^n}\big)^2\Big]} \\
&\leq& \sqrt{2^{n(Q - H(B) + H(E) + 3\d)}} \nn \\
&\leq& 2^{-n\d}\leq \ep \nn
\end{eqnarray}
provided that $\d \leq \smfrac 1c$.
By choosing $\d >0$ to be arbitrarily small, it therefore follows from Lemma~\ref{lem:decouple} and the above estimates that for every rate $0\leq Q \leq I_c(\ph^{A'},\CN)$, there exists a $(Q,n,6\ep)$ code for all sufficiently large $n$.  Furthermore, Markov's inequality implies that a randomly selected code will be a $\big(Q,n,\sqrt{6\ep}\big)$ code with probability at least $1-\sqrt{6\ep}$. 
\end{proof}

\section{Final remarks} \label{sec:epilogue}
We proved that the coherent information is an achievable rate for entanglement generation over a noisy quantum channel by showing that a state which is maximally entangled with a random subspace of the typical subspace is, with high probability, decoupled from the environment.  While this is not the first proof of the quantum channel coding theorem, by avoiding the need to explicitly construct and analyze the receiver's decoding operation, the proof becomes significantly simpler than other approaches. The observation that decoupling from the environment would be sufficient for quantum error correction goes back at least as far as the first analyses of the entropic conditions for quantum error correction~\cite{SN96}. Recently, versions of the idea as used here have been shown to have a wide range of applications in quantum information theory ranging from state merging and multiuser quantum data compression~\cite{HOW05b} to noisy channel simulation and entanglement-assisted communication over quantum channels~\cite{ADHW06}.

When proving coding theorems for network problems \cite{DS05,YDH05,YHD06}, it is often useful to start with randomized single-user codes whose average input is close to a product state \cite{D05}.  By absorbing the normalizations of the code states into the induced measure, the resulting expected input state to the channel is the normalized version of $\Pi_\d^{A'} (\ph^{A'})^{\ox n} \Pi_\d^{A'}$ which, in turn,  is close to $(\ph^{A'})^{\ox n}$ by typicality.  On the other hand, it is also possible to obtain codes which are maximally entangled with a uniformly random subspace of a subspace of the input by the following modification of our procedure.   Appendix~A of \cite{ADHW06} shows that for every $\delta>0$ and all sufficiently large $n$, there is a projection $\Pi_t^{A}$ acting on $A^n$ such that the normalized version $\ket{\ph_t}^{A^nB^nE^n}$ of the state $\Pi_t^A\ket{\ph}^{A^nB^nE^n}$ satisfies the following desirable properties.  First,  $\ph_t^{A^n}$ is maximally mixed on a subspace $A_t$ of dimension $\geq 2^{nH(A) - n\d}$.  Therefore, $\ket{\ph_t}^{A^nB^nE^n}$ can be obtained by acting with $\CU_\CN^{\ox n}$ on half of a suitable maximally entangled state.  Second, it is shown that this state is 
exponentially close (in $n\delta^2$) to a state $\ket{\ph_{t,\delta}}^{A^nB^nE^n}$ which satisfies the bounds (\ref{eqn:typ1}) and (\ref{eqn:typ2}).  Using these in place of the states $\ket{\ph}^{A^nB^nE^n}$ and $\ket{\ph_\d}^{A^nB^nE^n}$ introduced by Lemma~\ref{theo:typical} in the proof of Theorem~\ref{theo:maintheorem}, one finds that the random state $\ket{\psi}^{RB^nE^n}$ obtained by the covariant measurement on the $A_t$ subspace of $\ket{\ph_t}^{A^nB^nE^n}$ is always maximally mixed on $R$, and thus can be created by maximally entangling with a uniformly random subspace of $A_t$.

The random codes we have constructed use an infinite amount of common randomness, owing to the uncountability of the unitary group.  On the other hand, as argued in~\cite{ADHW06}, it is possible to replace the integrals over the unitary group with finite sums over any unitary 2-design, for which there are known efficient exact and approximate constructions~\cite{DLT02,dankert-2006}.
For example, the Clifford group forms a unitary 2-design, which implies that the coherent information can be achieved using random stabilizer codes~\cite{G97}. (The decoding procedure for these codes, however, need not be the standard stabilizer code decoding procedure.) This result was anticipated in work by Hamada, who showed that for a large class of input states, the coherent information can be achieved using stabilizer codes~\cite{hamada}.
In fact, it is possible to build on the results of the current paper to construct codes achieving the coherent information while being encodeable and decodable in polynomial time on a quantum computer~\cite{GHS06}.

For a general quantum channel, the best known expression of the quantum capacity is
\begin{equation}
\CQ(\CN) = \lim_{n\to\infty}\frac{1}{n}\CQ^{(1)}(\CN^{\ox n})
\label{eqn:capacity}
\end{equation}
where $\CQ^{(1)}(\CN) \equiv \max_{\ph^{A'}}I_c(\ph^{A'},\CN).$ This
follows by combining the coding theorem, which shows that
$\frac{1}{n}\CQ^{(1)}(\CN^{\ox n})$ is a \emph{lower bound} for the
quantum capacity, with a so-called \emph{multi-letter converse} (see
e.g.~\cite{D05}). The formula (\ref{eqn:capacity}) is of limited
practical use, however, as it does not seem to lead directly to a
computable expression for the quantum capacity. Indeed, it is
currently a major open problem to give an effective procedure for
computing the quantum capacity of an arbitrary quantum channel;  the
exact answer is not even known for the qubit $p$-depolarizing
channel $\CN_p(\rho) = (1-p)\rho + \frac{p}{2} \openone$. A notable
exception is the class of degradable channels, for which $\CQ^{(1)}$
is \emph{additive}~\cite{DS05}, meaning that $\CQ^{(1)}(\CN^{\ox n})
= n \CQ^{(1)}(\CN)$, which leads to a \emph{single-letter}
expression $\CQ(\CN) = \CQ^{(1)}(\CN)$ for the capacity of
degradable channels. This is known not to be the case for the
depolarizing channel~\cite{DSS98, SS07}. While additive upper bounds on
(\ref{eqn:capacity}) are known~\cite{SSW06},
it is entirely conceivable that a different coding strategy could
show the achievability of some other function on channels which
\emph{is} additive, leading to a complete characterization of the
quantum capacity as in the classical case \cite{shannon}.

\appendix 

\begin{proof}[Proof of Lemma~\ref{lem:decouple}]
We follow a line of reasoning similar to that introduced in \cite{sw-approx}.
For completeness, we give a version of the argument after recalling
some facts about distance measures. The fidelity can be defined for
an arbitrary pair of density matrices as $F(\rho,\s)
\equiv \Norm{\sqrt{\ph}\sqrt{\s}}_1^2$, where the \emph{trace norm}
$\norm{X}_1$ of an operator $X$ is the sum of its singular values.
Fidelity satisfies $0\leq F(\ph,\s)\leq 1$, where the second
inequality is saturated iff $\ph = \s$.  An alternate
characterization of fidelity, known as Uhlmann's theorem~\cite{U76},
says that given any purification $\ket{\ph}^{AB}$ of $\ph^A$, the
fidelity $F(\ph^A,\s^A)$ equals the maximum of
$|\braket{\ph}{\s}|^2$ over all purifications $\ket{\s}^{AB}$ of
$\s^A$.  The trace distance gives a lower bound to the fidelity (see e.g.\ \cite{nc}):
\begin{eqnarray}
F(\ph,\s) &\geq& 1-\norm{\ph - \s}_1 \label{tr2fid}
\end{eqnarray}
Furthermore, fidelity behaves monotonically under the action of a quantum channel $\CN$:
\begin{eqnarray}
F(\ph,\s) &\leq& F\big(\CN(\ph),\CN(\s)\big). \label{fidmon}
\end{eqnarray}
We now give the proof.  
By Uhlmann's theorem, there is a purification $\ket{{\Psi'}}^{RB^nE^n}$ of $\pi^R\otimes \ph^{E^n}$ satisfying
\[|\braket{\Psi}{\Psi'}|^2 = F\big(\Psi^{RE^n},\pi^{R}\otimes\ph^{E^n}\big).\]
Because $\Psi'^R=\pi^R$ is maximally mixed, it is also purified by a maximally entangled state $\ket{\Phi}^{R\h{R}}$.
Furthermore, since $\pi^R\otimes \ph^{E^n}$ is a product state, it must have a purification which is a tensor product of pure states.  Therefore, there is another Hilbert space $B'$, a pure state $\ket{\xi}^{B'E^n}$ and an isometry $\CW^{B^n \to \wh{R} B'}$ under which
\[\CW^{B^n\to \wh{R}B'}\ket{\Psi'}^{RB^nE^n}  =  \ket{\Phi}^{R\h{R}}\ket{\xi}^{B'E^n}.\]
Combining monotonicity of fidelity (\ref{fidmon}) with the relation (\ref{tr2fid}), this implies that the decoding $\CD^{B^n\to \h{R}}\equiv \Tr_{B'}\CW^{B^n\to \h{R}B'}$ satisfies (\ref{eqn:th2fidelity}) as required.
\end{proof}

\begin{proof}[Proof of Lemma~\ref{lem:trdist}]
By the triangle inequality, 
\begin{eqnarray*}
\norm{\rho - \sig}_1 &\leq&  \norm{\rho - c\rho}_1 + \norm{c\rho - \sig}_1 \\
&=& |1-c| + \norm{c\rho - \sig}_1.
\end{eqnarray*}
However, $|1-c| = \big|\!\Tr\,[c\rho - \sig]\big| \leq \norm{c\rho - \sig}_1$ and the lemma is proved. 
\end{proof}

\begin{proof}[Proof of Lemma~\ref{lem:twirl}]
Observe that $G$ is invariant under the representation $U\ox U$ of the unitary group whose restrictions to the symmetric and antisymmetric subspaces of $A\ox A$ are irreducible.   Writing $\Pi^A_\pm$ for the projections onto these respective subspaces, Schur's lemma implies that $G$ has the form 
\begin{eqnarray*}
G &=& \smfrac{1}{D_+} \Tr\big[F^R \Pi^A_+\big] \Pi^A_+ + \smfrac{1}{D_-}\Tr\big[F^R \Pi^A_-\big]\Pi_-
\end{eqnarray*}
where $D_\pm = \smfrac 12(D^2 \pm D) = \Tr\,\Pi_\pm^A$. Writing $\Pi_\pm^R$ for the projections onto the symmetric and antisymmetric subspaces of $R\ox R$, observe that we may write each of the flip operators as $F^{\{A,R\}} = \smfrac 12 \Pi_+^{\{A,R\}} - \smfrac 12 \Pi_-^{\{A,R\}}$.  Since $\Pi^R_\pm \Pi^A_\pm = \Pi^R_\pm$ and $\Pi_\pm^A\Pi_\mp^R = 0$ we have, for $\Tr\,\Pi^R_\pm =\smfrac 12(d^2\pm d) \equiv d_\pm$
\begin{eqnarray*}
G &=& \smfrac{d_+}{D_+} \Pi^A_+ + \smfrac{d_-}{D_-} \Pi_-^A \\
&=& \smfrac 12\!\left(\smfrac{d_+}{D_+} +  \smfrac{d_-}{D_-}\right)\openone^{AA} + 
 \smfrac 12\!\left(\smfrac{d_+}{D_+} +  \smfrac{d_-}{D_-}\right)F^A
 \end{eqnarray*}
 as required.
\end{proof}

\begin{proof}[Proof of Lemma~\ref{lem:covariant}]
The trace norm of a Hermitian operator $X$ can be expressed as $\norm{X}_1 = \max\{ \Tr\, Y X : -\openone \leq Y \leq \openone\}$.
Suppose now that the random matrix $Y_V$ achieves the maximum for $\norm{VXV^\dagger}_1$, so that   $-\openone \leq Y_V \leq \openone$ and  $\norm{VXV^\dagger}_1 = \Tr Y_V V X V^\dagger$. In particular, observe that $-V^\dagger V \leq V^\dagger Y_V V\leq V^\dagger V$.  Together with our assumption $\bE[V^\dagger V]\leq \openone$, this yields the inequalities  $-\openone \leq \bE[V^\dagger Y_V V] \leq \openone$.  We therefore conclude from cyclicity of the trace and linearity of expectation that
\[\bE\Norm{VXV^\dagger}_1 = \Tr\,\bE[V^\dagger Y_V V] X  \leq \norm{X}_1.\]
\end{proof}
\acknowledgements
We would like to thank Alexander Holevo, Debbie Leung, Renato Renner
for interesting discussions. PH is supported by the Canada Research
Chairs program, CIFAR, FQRNT, MITACS, NSERC and QuantumWorks.  He is also grateful
to the DAMTP in Cambridge for their hospitality.  MH is
supported by Polish Ministry of Scientific Research and Information
Technology under the (solicited) grant no. PBZ-MIN-008/P03/2003 and
EC IP SCALA.
AW was supported through an Advanced 
Research Fellowship of the U.K.\ EPSRC via the ÒQIP IRCÓ, and the European Commission 
IP ÒQAPÓ (IST-2005-015848).  JY's research at LANL is supported by the Center for Nonlinear Studies (CNLS), the Quantum Institute and the LDRD program of the U.S.\ Department of Energy.  He is also grateful for support from the U.S.\ National Science Foundation under Grant No.\ PHY-0456720 through Caltech and from the CIFAR when this work was initiated. 

\theendnotes

\bibliographystyle{klunum}
\bibliography{simple}
\end{document}